# Marine Buoy Location Finding and Tracking System for Linux Supporting Mobiles

Harikrishnan.R, Shajna S. Hammed, P.Malini
*Department of Electrical and Electronics Engineering*
*Vel Tech Dr.RR & Dr.SR Technical University Avadi Chennai-600062*

*Abstract* - **Marine buoy is an important part of underwater acoustic communication system. It is of great significance to track and locate it. It is widely used in ocean environment three-dimensional monitoring, underwater multimedia communication, underwater mobile carrier navigation and positioning, marine resources detection, remote control of submarine topography mapping and offshore oil industry, data acquisition, etc. This paper describes the application of the monitoring service of GPRS / GPS module at Marine buoy. It can achieve real-time location of underwater acoustic communication devices and route tracking to avoid the loss of the device, as well as assist to retrieve the lost device.**

## I. INTRODUCTION

The essential difference between underwater acoustic communication and wireless communication is the different physical characteristics of their propagation mediums. Due to the characteristics of high consume of the radio in the water, it is unrealistic to use the land fairly mature wireless communication technology in underwater communication. Underwater cable less communication has been the obstacle of "sea, land and air" three-dimensional interconnection for a long time. So a lot of maturity program of the wireless communication cannot be directly applied in the underwater acoustic communication. Establishing the oceanographic survey platform to obtain cumulative data analysis will help to study the channel's regularity. What's more, it can provide a relatively convenient environment to verify the feasibility of the simulation results in a real environment and facilitate the research work carried out smoothly.

The sea buoy system, as a carrying platform of launch, acquisition and GPS and other equipments, connects with the control center via the wireless network to transmit data and receive instructions from the control center. The buoys equipment communicates with each other through the underwater acoustic channel. The platform is convenient for experiments carried on the sea, and enables remote monitoring and management.

The control center is mainly responsible for the buoy system GPS information data receiving, extraction, processing. It also reads and writes data records to the database, transmits data information to the JavaScript server, and receives feedback information, and finally processes and displays the information. JavaScript server plays a role in receiving terminal data and communicates with Baidu Maps API service to achieve related functions.

Oracle database provides data services. Sea buoy as an important part of the maritime communication system has an important place in the underwater acoustic communication. So it is essential to track and locate it.

**Website System**

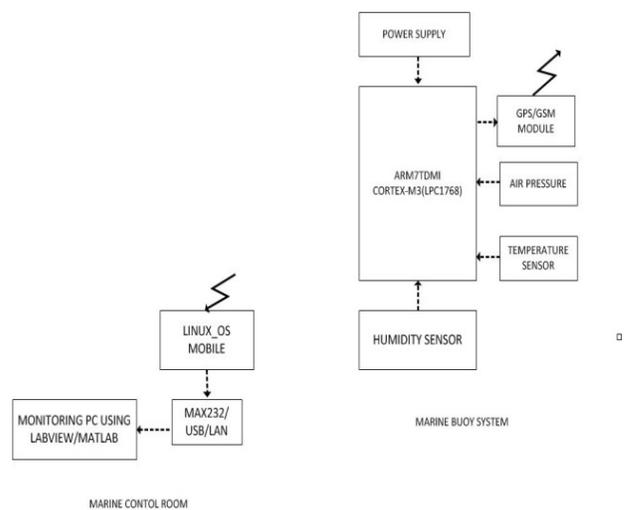

Fig.1 Block Diagram of the Prototype Buoy System for Wireless System

## II. THE DESIGN OF THE LOCATOR

GPS is currently the most widely used positioning system and the technical is also very mature. Due to the special circumstances of the marine environment, wired communication by a large degree of restrictions is obviously unrealistic. With the consideration of the wide coverage of the GSM network and the stable signal, the GPS satellite positioning and GPRS data transmission function perfectly integration will be a good design. The paper will discuss the application of highly integrated GPRS / GPS tracking locator in sea buoy tracking and positioning. The design of the locator includes both hardware and software design.

### A. Hardware Design
We use the GSM and GPS chip as the design module. The module can be widely used in a variety of positioning fields with its expansion application as remote positioning terminal. GSM baseband chip is a highly integrated chip PMB7880 of Infineon. It has an advantage of low price and small size [1]. GPS positioning chips is the third generation satellite positioning receive chip SiRF star III of SiRF[2] can simultaneously track 20 satellite channels, high positioning accuracy, high sensitivity and fast positioning.

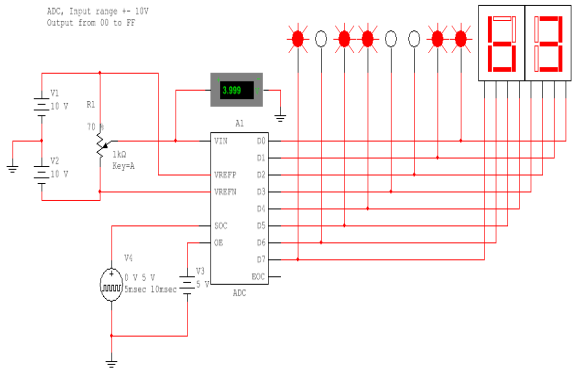

Fig 2a Modular structure diagram

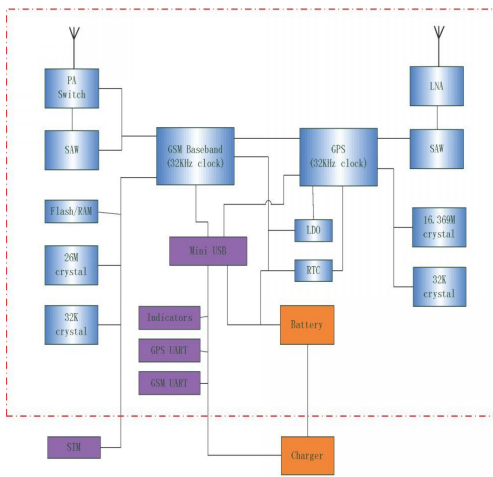

Fig 2b Modular Framework diagram

*i) The Module Diagram*

The terminal product is function expansion based on the module, such as adding SIM card, antenna, speaker and so on. The locator's block diagram is shown below

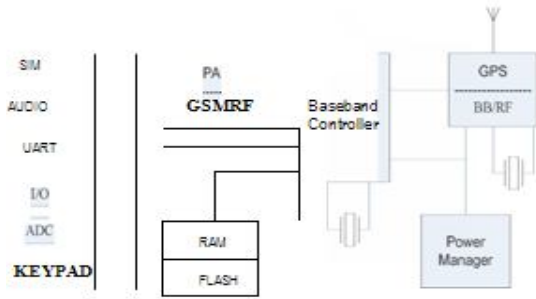

Fig 2c Design Framework

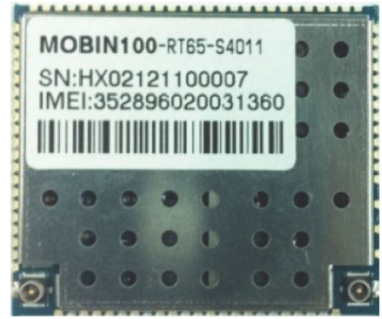

Fig 2d Module Appearance

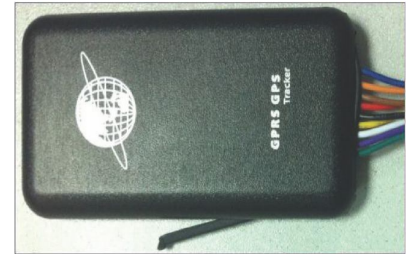

Fig 2e Locator Appearance

*B. Software design*

The module will call Baidu Maps API method through the JavaScript technology [3] to achieve the following map service functions: real-time monitoring, track playback, electronic fences, map labels, and other functions. JavaScript and Oracle [4] connect through OLEDB. We can get history records, latitude and longitude information extracted from the records and playback track through connection between JavaScript and the database. Since the received data of the GPS receiver contains a variety of data and in fact we are concerned only to the $GPRMC beginning data block [5], we have to process it into the information which is easily understood and convenient to be displayed.

The locator will register mobile network, connect GPRS, open GPS positioning, open analog serial to analyze GPS data and lit indicator after start. It will turn off GPS and GPRS in a power-saving mode when there is no data request.

The terminal locator uses GPRS of SMS to communicate with the server. GPRS communication is essentially based on TCP/IP protocol TCP for data transmission. It uses request / response synchronous communication model. When using SMS, SMS encodes with binary PDU. Figure 4 shows the structure of the protocol stack.

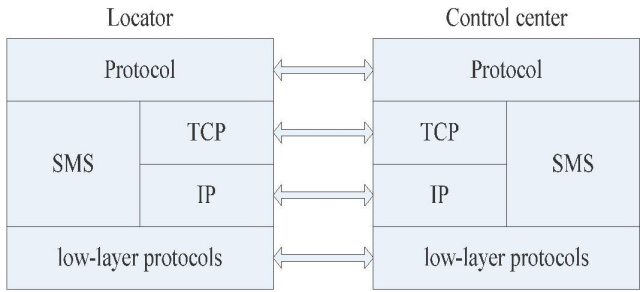

Fig 2f Mobile Positioning Communication Protocol Stack Structure

### III. EXPERIMENT RESULT

The control center can view the real-time location information of the monitored equipment, and display on map of Baidu. The locator's major functions are as follows:

1. Real-time monitoring: Real-time displays of actual position information of online terminals without refresh the page.

2. Track playback: It's convenient for users to query trajectory of the terminal equipment within a certain period of time. You just need to select the terminal name or IMEI number, set the starting and ending time and playback speed, and then can choose to play (Track playback time cannot be setted more than 7 * 24 hours).

3. Electronic fence: According to the users' needs to set an arbitrary polygon and polygon set for the security of the terminal. If the terminal position is out of the range, it will send alarm information to the host.

4. Map labels: Users can make text label on the position they interested. The text annotation can be saved or deleted.

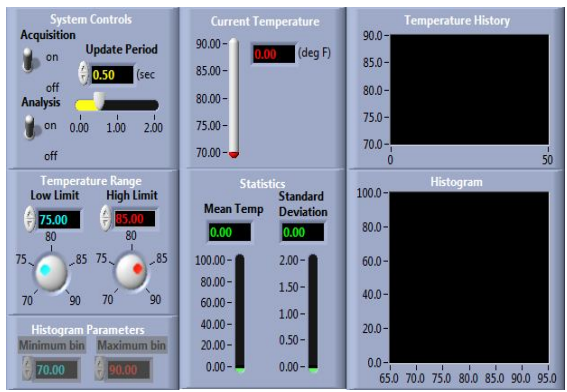

Fig.3a Software Result

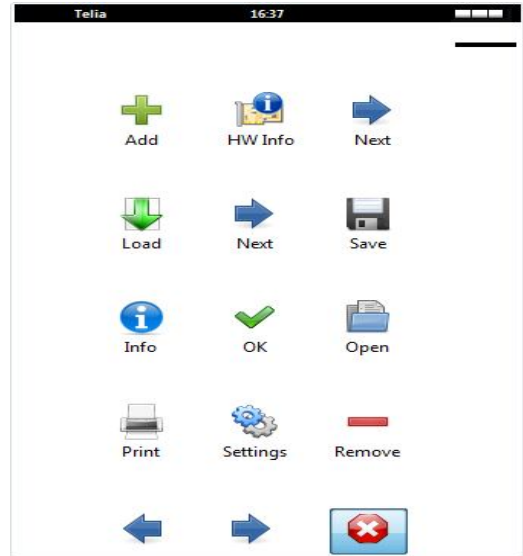

Fig.3b Map Labels

### IV. CONCLUSION

GPS satellite positioning and GPRS data transmission functions perfectly integrated locator used for tracking and positioning of the buoys can facilitate the research work, and furthermore it reduces the expensive the possibility of underwater acoustic communications equipment's' accident lost or stolen.

### ACKNOWLEDGMENT

This work was supported by the Fundamental Research Funds for the Central Universities (2011121050, 2012121028), the National Natural Science Foundation of China (61001142, 61071150) and the Science Technology Project of Xiamen Government (3502Z20123011)